# Growth kinetics of a nanoprotuberance under the action of an oscillating nanotip

J. P. Aimé,* D. Michel, R. Boisgard, and L. Nony
*CPMOH, Université Bordeaux I, 351 Cours de la Libération, F-33405 Talence Cedex, France*

The atomic force microscope is a versatile tool that allows many routes to be used for investigating the mechanical properties of soft materials on the nanometer scale. In the present work, experiments were performed on polystyrene polymer films of various molecular weight by approaching a vibrating nanotip towards the surface. The variation of the oscillating amplitude of the cantilever is interpreted as the result of the growth process of a nanoprotuberance. The growth rate is found to be dependent of the magnitude of the oscillating amplitude and of the molecular weight. A model is developed describing in a very simple way the action of the tip and a viscoelastic response of the polymer. The numerical simulation helps in understanding the nonlinear relation between the growth rate and the vibrating amplitude of the microlever and describes qualitatively most of the experimental features. For the softer material, experimental situations are found that allow the experimental results to be amenable with an analytical solution. The analytical solution provides a fruitful comparison with the experimental results showing that some of the nanoprotuberance evolution cannot be explained with the approximation used. The presents results show that there exists a new and fascinating route to better understand the mechanical response at the local scale. 

## I. INTRODUCTION

The atomic force-microscope (AFM) is frequently used to investigate surface properties through the study of the oscillating behavior of a cantilever. The mechanical vibration of a cantilever was first used to provide an image of the force gradient variations above the surface. A linear analysis shows that force gradients are detected as shifts in the resonant frequency.[1] Taking advantage of the tip-surface interaction, this mode of analysis has a variety of applications, including noncontact surface profilometry,[2,3] imaging localized charge,[4,5] and, recently, providing a contrast at the atomic scale.[6–9] Nevertheless, when the tip is very close or slightly touches the surface the system is highly nonlinear.[10–13] This nonlinear behavior increases the complexity of the cantilever response, and the physical origin of the contrast at the atomic scale becomes more difficult to describe, but in turn the unstable behavior of the cantilever can be used to improve the sensitivity of the AFM.

In Ref. 13 it was shown that the attractive interaction between the tip and the surface is able to strongly modify the oscillating behavior of the tip-cantilever system. As soon as the tip is near the surface, typically for a tip-surface distance of approximately 1 nm and for a drive frequency slightly below the resonance frequency, the oscillator shows a bifurcation from a monostable to a bistable state. The bifurcation leads to a cycle of hysteresis. The attractive force field introduces nonlinear coupling terms that make the magnitude of the drive amplitude a key parameter. An analytical expression is derived showing that the shape of the hysteresis is dependant of the drive amplitude through a cubic law.[13] A well-defined cycle of hysteresis can be experimentally obtained, but this happens strictly for surfaces that are inert mechanically, or that have a local stiffness large enough for an elastic displacement not to be induced.

The scope of the present work is to investigate what hap- P pens when a sample has a mechanical susceptibility, the reciprocal of a stiffness, that allows the surface to exhibit a sizable elastic displacement under the action of the tip. The aim is to explore a new route to probe the mechanical response of a surface at the nanometer scale without touching or only slightly touching the surface.

The paper is organized as follows, in a second section we detail the experimental methodology and give the experimental results. In the third section the main ideas of the theoretical simulation that describes the kinetics of a nanoprotuberance are introduced. In the last section we discuss the usefulness and the limits of the hypothesis that sustain the simulation and introduce a simple rheological model. The phenomenological model is essentially heuristic and should help us to define new experiments and further quantitative analysis.

## II. EXPERIMENTAL SECTION

### A. Methodology

The experiments have been performed in air and recordings were made by approaching the surface towards the tip using the tapping mode of a nanoscope III.[14] Two piezoactuators are needed to perform an experiment in the tapping mode. A small piezo allowing the microlever to be vibrated at a frequency close to its resonance frequency and a second piezo moving the sample. The piezoactuator holding the sample is a piezo with a maximum vertical displacement of 660 nm giving a low noise signal with a good resolution.

At a fix position of the plane surface, the sample is approached towards the tip. It then is retracted when the tip is near the surface. To get a simple criterion, the drive frequency is slightly below the resonance frequency, such that the vibrating amplitude is below the resonant one. With this condition, the amplitude of the vibrating cantilever increases when the attractive force field between the tip and the surface becomes large enough. Therefore, as soon as an increase of the amplitude is observed, the sample is retracted and the range of the tip-sample distance is restricted to the domain represented by the circle drawn in Fig. 1.

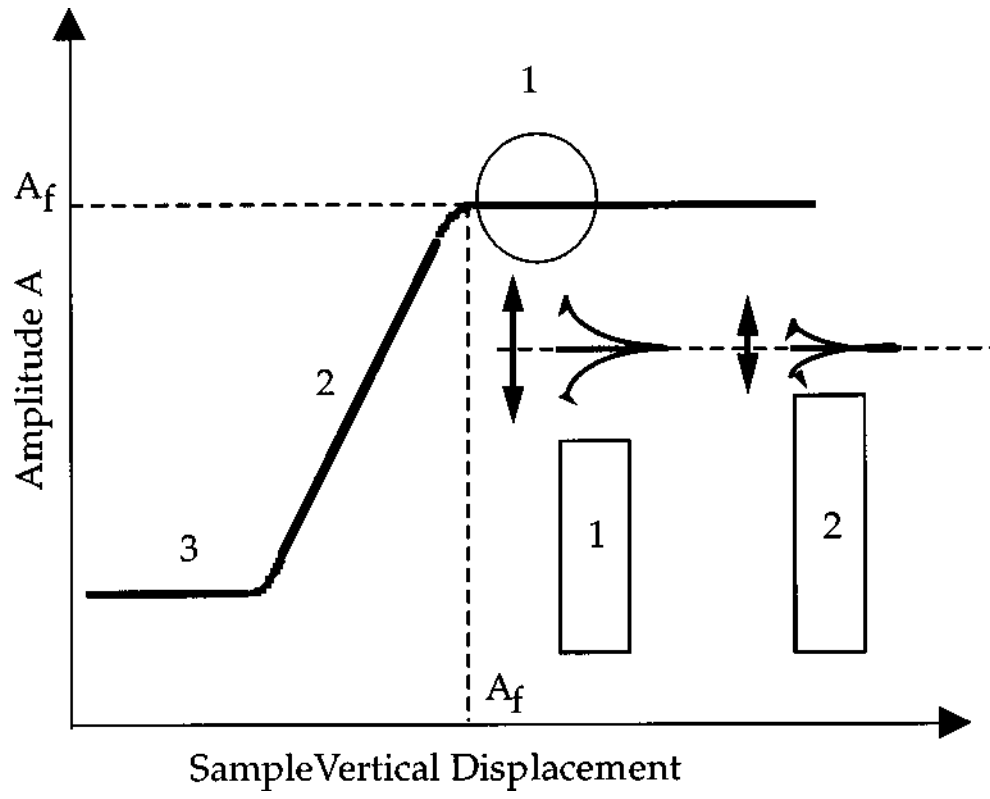

FIG. 1. Typical *S* shape of the oscillations as a function of the distance between the surface and the position at rest of the microlever. $A_f$ is the amplitude of the oscillations far from the surface, typically for distance above 20 nm$+A_f$. For infinitely hard surface, the amplitude decreases linearly with the distance as soon as an intermittent contact happens (domain 2). When the distance between the equilibrium position at rest of the microlever and the surface is zero, the amplitude becomes zero (domain 3). Experiments have been performed in the domain defined by the circle corresponding to distances larger than $A_f$ (domain 1).

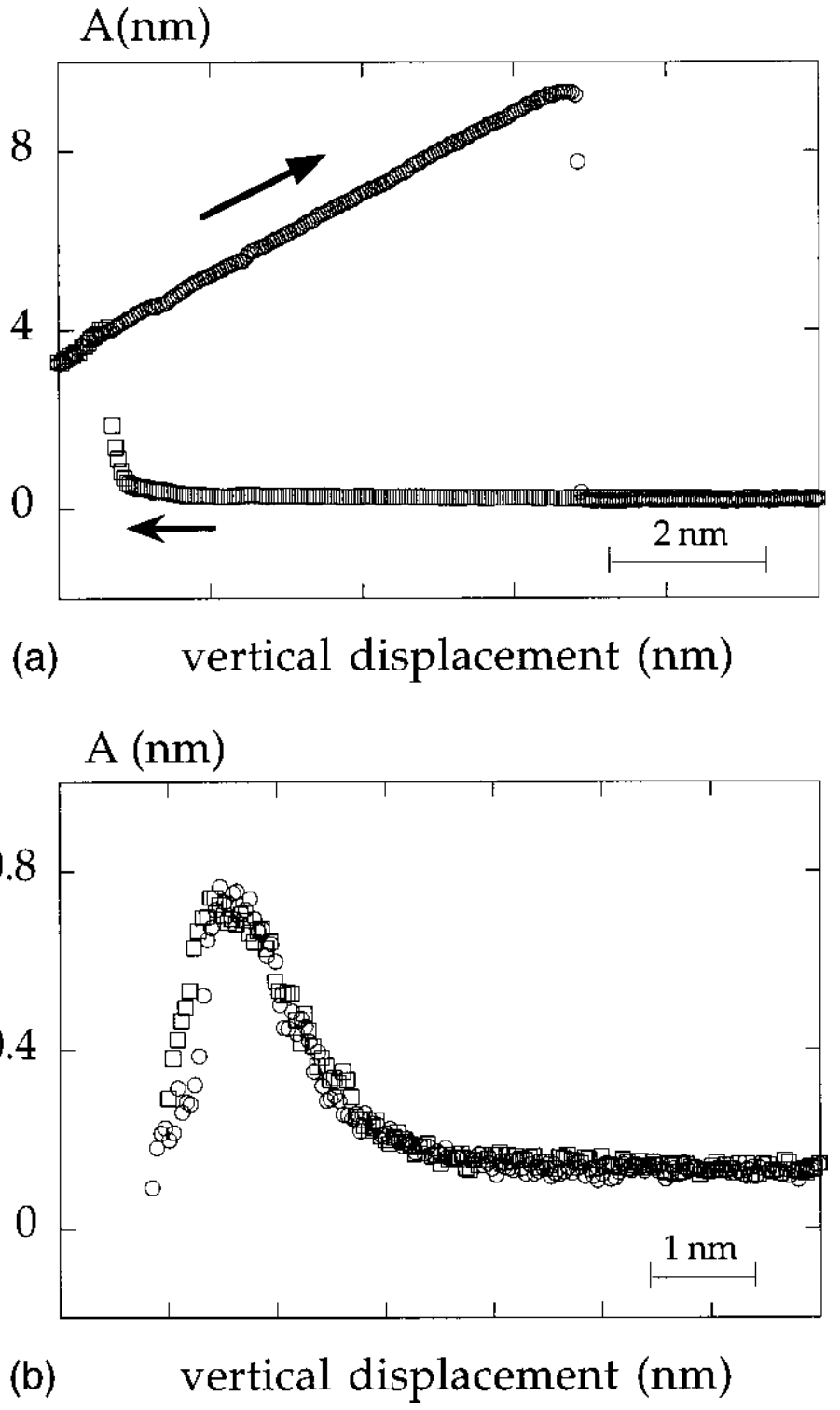

FIG. 2. The usual approach-retracting curve giving the variation of the oscillating amplitude as a function of the vertical displacement of the surface. The sample displacement is performed with a scan size of 20 nm at a scan frequency of 10 Hz. The empty square and empty circle correspond to the approach and the retraction motion of the sample. (a) $A_f=44$ nm, (b) $A_f=2$ nm. The sample is the polymer film of $M_w=150\,000$.

Far from the surface, the vibrating amplitude of the cantilever, here after noted the free amplitude $A_f$, is given by the well-known expression

$$A_f(\omega)=\frac{a\,\omega_r^2}{\sqrt{(\omega_r^2-\omega^2)^2+(2\beta\omega)^2}}, \tag{1}$$

where $a$ and $\omega$ are the drive amplitude and the drive frequency of the small piezo, $\omega_r$ and $\beta$ the resonance frequency and the damping factor of the cantilever.

The variation of the amplitude as a function of the tip-surface distance depends on the magnitude of the free amplitude. For large free amplitudes $A_f$ a cycle of hysteresis follows, while for small amplitudes $A_f$ the same variation is observed during the approach and the retraction of the sample (Fig. 2). In other words, for small $A_f$, whatever the distance between the tip and the sample, the variation of the amplitude is reversible, while a cycle of hysteresis is observed for large $A_f$ values. Theoretically, the whole features can be described as a function of the magnitude of the free amplitude.[13] At small $A_f$ values, the effect of the nonlinear coupling terms vanishes and the oscillating behavior of the cantilever is satisfactorily described by a linear analysis. Thus, for a surface that is mechanically inert, a typical response of the tip-cantilever system will be identical either to the one shown in Fig. 2(a) or in 2(b).

The samples are polystyrene (PS) films. Several PS films have been prepared with molecular weights varying between 284 000 and 1890 and polydispersities ranging between 1.03 and 1.07 [Fluka, Chemika, Buchs/Switzerland]. The baseline idea is that an organic material interacts with the tip uniquely through the van der Waals London dispersive force so that the strength of the attractive interaction is simply given by the product of the radius of the tip and the Hamaker constant [see Eq. (4)]. The Hamaker constant is a function of the chemical composition and of the density of the two interacting materials,[15] thus does not depend of the molecular weight. Consequently, by using the same tip to investigate samples of various molecular weights, thus keeping the same radius of curvature of the tip, the strength of the attractive interaction between the tip and the surface is a constant independent of the molecular weight.

Besides, at the local scale, the mechanical properties of the surface were shown to be molecular weight dependent.[16] The smaller the molecular weight, the softer the material. Therefore, if any change of the oscillating behavior occurs as a function of the PS films properties, one expects that it will correspond to the change of the mechanical properties of the PS films and not to the change of the tip-PS attractive interaction.

### B. Experimental results

The resonance frequency of the cantilever is $\nu_r=293.23$ kHz, the quality factor is $Q=450$. The damping factor of the damped oscillator is $\beta=\nu_r/2Q$. Most of the data reported were obtained with a scan size of 20 nm and a scan frequency of 10 Hz; a few of them were obtained with a scan frequency of 0.5 Hz. The drive frequency $\nu = 292.57$ kHz is kept the same throughout the experiments.

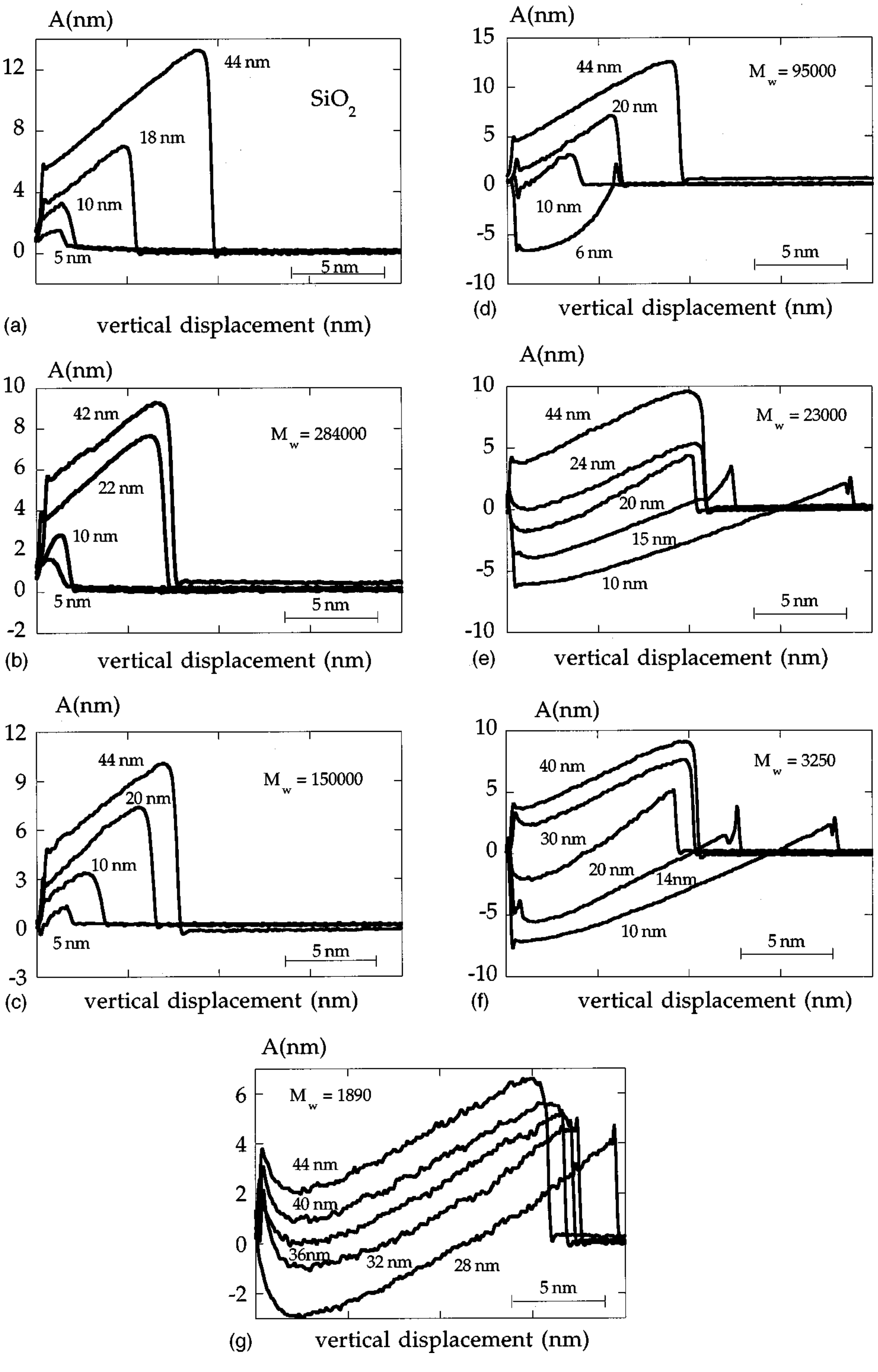

FIG. 3. Variations of the amplitude of the oscillation during the retraction of the surface as a function of the free amplitudes $A_f$ for different samples. Scan size 20 nm, scan frequency 10 Hz.

For the sake of clarity, we report selected experimental data corresponding to the retracting displacement of the sample. These data have been selected from among 10–20 experiments for each sample, with $A_f$ varying from 5 up to 44 nm. To be more easily compared, the experimental data have been subtracted from their $A_f$ values (Fig. 3).

In Figs. 3(a)–3(c) are reported the variation of the oscillating amplitudes obtained with a silica surface and the PS films of high molecular weights, respectively, $M_w = 284\,000$ and $M_w = 150\,000$. The silica surface has been prepared as described in Ref. 17. The silica surface is used as a hard reference surface from which we do not expect an elastic response. The three surfaces give similar variations of the amplitude as a function of the vertical displacement of the sample. Irrespective of the magnitude of the free amplitude, down to $A_f = 5$ nm, the experimental results are in good agreement with the theoretical predictions and the whole variations can be ascribed with the effect of the attractive tip-surface interaction.[13]

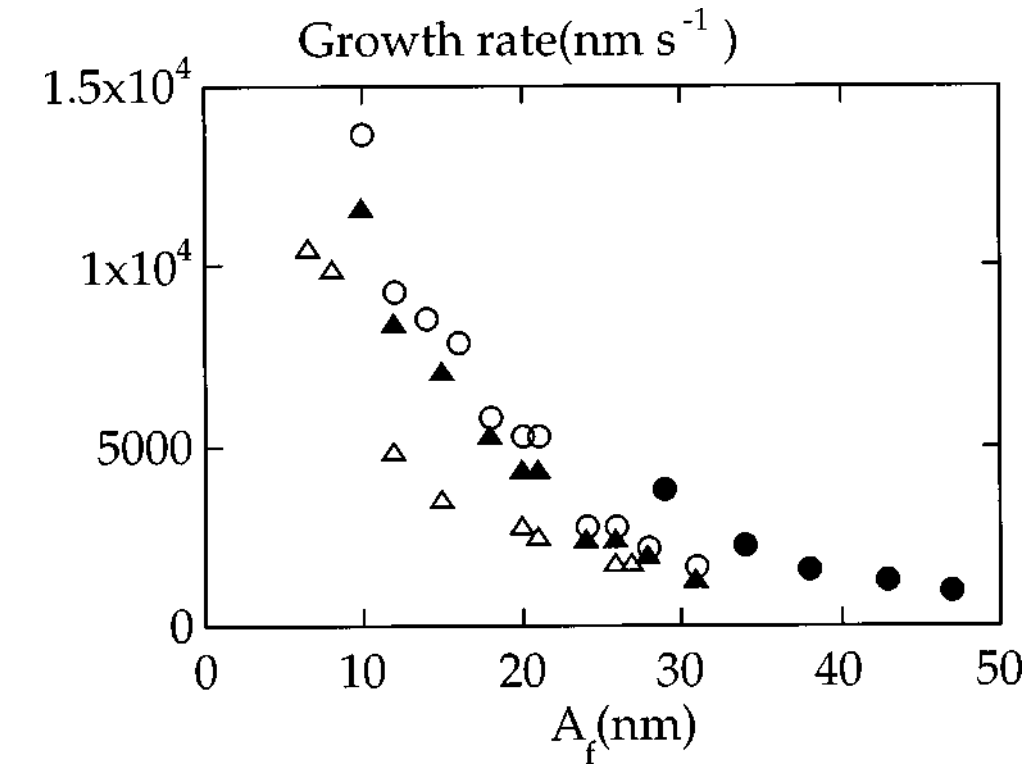

FIG. 4. Growth rates of the nanoprotuberance as a function of the starting amplitude. The growth rates are given by the fitted straight line at the beginning of the abrupt decrease of the oscillating amplitude. Empty triangle $M_w = 95\,000$, filled triangle $M_w = 23\,000$, empty circle $M_w = 3250$, filled circle $M_w = 1890$.

For the sample of molecular weight $M_w = 95\,000$ a drastic change is observed at the small $A_f$ values 10 and 6 nm [Fig. 3(d)]. In place of the usual loop, the amplitudes abruptly decrease, almost down to a zero value. One can also note a variation of the amplitude showing a slight perturbation at the beginning of the cycle for the intermediate values of $A_f$.

The response of the vibrating cantilever shows a completely different behavior for the samples of $M_w = 23\,000$ and $M_w = 3250$ [Figs. 3(e) and 3(f)]. Except for the highest $A_f$ values, the cycles of the hysteresis show a marked decrease of the amplitude just after the retracting displacement has been started. The smaller the free amplitude, the more pronounced is the minimum of the amplitude. For $A_f = 10$ nm, the amplitude decreases by about 6 and 7 nm, respectively, for $M_w = 23\,000$ and $M_w = 3250$.

The results show that the oscillating behavior cannot be described by uniquely considering the attractive force field. For example, let us compare the results obtained with the polymer of $M_w = 3250$. For $A_f = 42$ nm, a normal cycle of the hysteresis is observed and the amplitude returns to its $A_f$ value after a sample displacement of 10 nm has been made. For a smaller free amplitude, we normally expect a smaller width of the hysteresis as observed for higher molecular weights and the silica surface (see Figs. 3–5). Therefore the oscillator should recover its free amplitude value at a sample displacement smaller than 10 nm. The opposite behavior is observed, at $A_f = 14$ nm, the oscillator goes back to its $A_f$ value after a sample displacement of 13 nm. Such a behavior is even more pronounced for $A_f = 10$ nm. Similar observations, can be done for the sample of $M_w = 23\,000$.

For the lowest molecular weight investigated, $M_w = 1890$, we were unable, at least with a vertical scan size of 20 nm, to get a complete cycle with a free amplitude $A_f$ smaller than 28 nm [Fig. 3(g)].

In order to better understand what is really taking place, we focus on the abrupt decrease of the amplitude at the beginning of the retracting displacement. At the beginning of the abrupt decrease of the amplitudes, slopes can be extracted and multiplied by the piezoactuator velocity to give the rates at which the amplitude varies. In Fig. 4 are reported the measured rate of the variation of the oscillating amplitudes as a function of $A_f$ for four molecular weights.

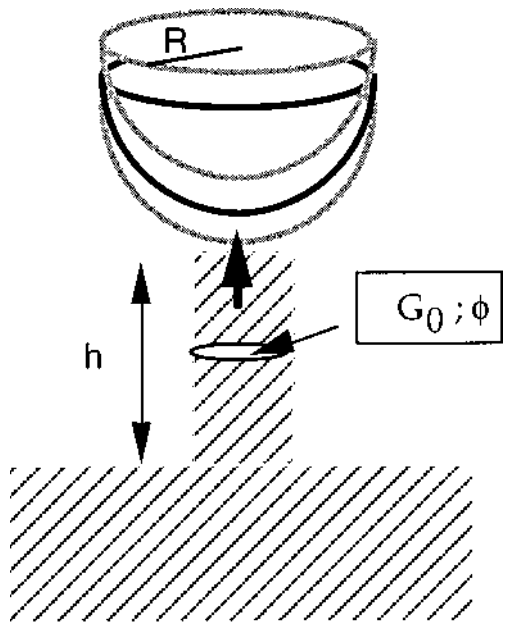

FIG. 5. Sketch of a nanoprotuberance created under the action of a oscillating nanotip. The geometry of the nanoprotuberance is the simplest one: a cylinder of diameter $\phi$ and height *h*. $G_0$ is the associated elastic modulus. *R* is the radius of the tip.

A general trend is easily extracted. The rate at which the amplitudes vary shows a nonlinear dependence on $A_f$ and increases as $A_f$ decreases. Furthermore, from the highest to the lowest molecular weight, the velocity shows an increased sensitivity to the magnitude of $A_f$. For $M_w$ higher than 95 000, as noted above (Fig. 3), no noticeable decreases of the amplitude were measured allowing a measurement of the slope to be reported. Conversely, for the low molecular weight $M_w = 1890$, even at a high value of $A_f$ a measurable decrease is observed. For example, the same velocity is measured for $A_f = 25$ nm ($M_w = 23\,000$) and $A_f = 35$ nm ($M_w = 1890$).

To interpret these evolutions requires consideration of an additional interaction in which the sample properties, apart from the Hamaker constant, must be included. Considering what is known about the intermittent contact,[10,19–21] we assume that the tip touches the sample and that the true vertical position of the sample is not the one given by the vertical displacement of the piezoactuator holding the polymer film. In other words, we assume a local elastic response creating a nanoprotuberance such that the tip touches the surface at a vertical position higher than the one monitored by the piezoactuator. Such a situation is drawn in Fig. 5.

## III. SIMULATION OF A LOCAL VISCOELASTIC RESPONSE UNDER THE ACTION OF THE TIP

### A. Preliminary remarks

It is worthwhile to examine first how a vibrating cantilever measures the growth of a nanoprotuberance. As the nanoprotuberance can grow rather quickly, the experimental results give velocities as large as 14 000 nm/s, one has to verify the ability of the oscillator to measure such a rate of perturbation.

Assuming that the variation of the oscillating amplitude describes properly the growth process means that the oscillator adiabatically responds to the perturbation. Let us note $h(t)$, the height of the elastic displacement at time *t*. If the oscillator follows exactly the change of the height the decrease of amplitude is given by

$$A(t) = A(t=0) - h(t). \tag{2}$$

Equation (2) is true if and only if the perturbing process is slow in comparison to the dissipative processes leading to the establishment of equilibrium in the oscillator. The adiabatic condition writes

$$dh/dt \ll A(t=0)/\tau \tag{3}$$

for a resonance frequency of 293.23 kHz and a quality factor of 450, the relaxation time of the oscillator is $\beta^{-1}=\tau=4.9\times 10^{-4}$ s. For $A=10$ nm, this leads to a maximum velocity $A/\tau=2\times 10^4$ nm/s. Thus a growth rate of $10^4$ nm/s will not be measurable. Indeed, for intermittent contacts, as happens for nonlinear phenomena, one can demonstrates rigorously that the adiabatic criterion is not given by Eq. (3). In the nonlinear regime, $\tau$ must be replaced by a value close to the oscillating period, here $T=3.4\ \mu$s. The physical reason is that the fluctuation-dissipation theorem, from which the damping coefficient $\beta$ of the oscillator is calculated, does not apply when a nonlinear response must be considered. Therefore the order of magnitude of the maximum velocity is now better given by $A/T=3\times 10^6$ nm/s, making a growth rate of $10^4$ nm/s measurable.

We shall proceed as follows: (i) In a first step we describe the action of the tip on the polymer. A very simple approach is employed, the aim being not to get a quantitative agreement with the experimental data, but to capture most of the physical phenomena. (ii) In a second step a possible mechanical response of the polymer is given.

### B. Description of the tip-surface interaction

The attractive interaction between a sphere and a plane surface is used as the external force acting on the polymer:[15]

$$F_{\text{ext}}=\frac{HR}{6d^2}, \tag{4}$$

where $H$ is the Hamaker constant, $R$ the radius of curvature of the tip, and $d$ the distance between the tip and the sample. Taking the case where the distance between the surface and the equilibrium position of the cantilever at rest is equal to the free amplitude $A_f$, one gets the time dependence for the force:

$$F_{\text{ext}}(t)=\frac{HR}{6[A_f+0.165-A_f\cos(\omega t)]^2}, \tag{5}$$

where $\omega$ is the drive frequency and 0.165 nm is the ''contact'' distance for most of organic materials.[15] The contact distance is used in Eq. (4) to eliminate the diverging behavior. Let's now define a distance $d_c$ between the tip and the surface above which the action of the force becomes negligible. Typically, with $H=5\times 10^{-20}$ J and $R=10$ nm, for $d$ between 1 and 0.165 nm, $F_{\text{ext}}$ varies between 0.1 and 3 nN. The approximation showing that the influence of the tip is negligible for values of $d$ above $d_c$ is

$$A_f[1-\cos(\omega t)]\leq d_c,$$

which leads to the definition of a maximum residence time of the tip near the surface given by

$$\tau_{\text{res}}=\frac{T}{\pi}\cos^{-1}\left(1-\frac{d_c}{A_f}\right). \tag{6}$$

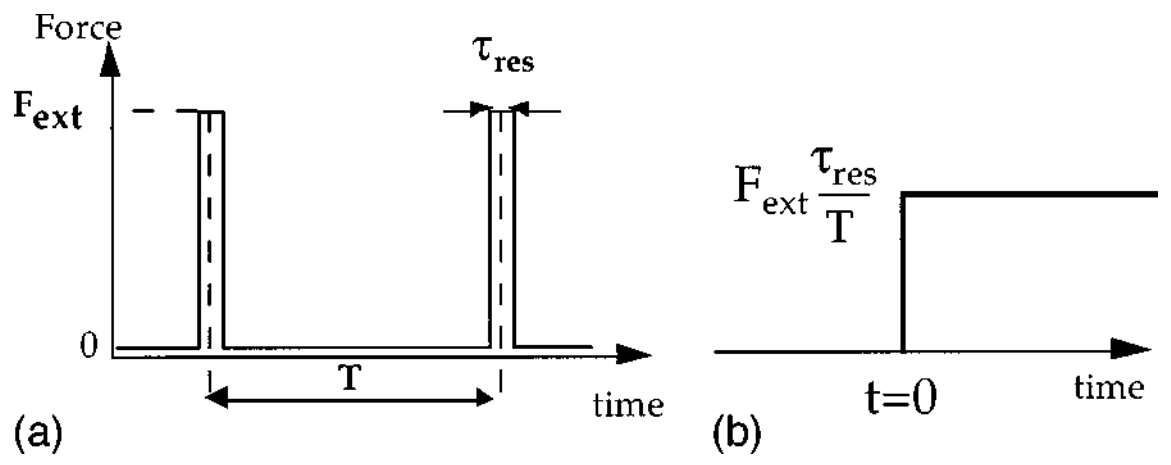

FIG. 6. (a) Periodic rectangular function describing the action of the tip. $T=3.4\times 10^6$ s, the width $\tau_{\text{res}}$ of the rectangular function is given by Eq. (6). (b) Sketch of the action of the oscillating tip with the assumption that the static part is the leading term [Eq. (7)]. The height of the step function is varying with time (see Appendix).

The larger the amplitude $A_f$, the shorter the residence time $\tau_{\text{res}}$.

A further approximation is to consider a periodic rectangular function of width $\tau_{\text{res}}$ and period $T=2\pi/\omega$ [Fig. 6(a)]. The Fourier transform of the rectangular function is

$$F_{\text{ext}}(t)=F_{\text{ext}}\left(\frac{\tau_{\text{res}}}{T}+\sum\ \frac{1}{\pi n}\sin(n\omega\tau_{\text{res}})\cos(n\omega t)\right). \tag{7}$$

The static part of $F_{\text{ext}}(t)$ is proportional to $\tau_{\text{res}}$, which in turn increases as $A_f$ decreases. Therefore, even within this rough approximation, the main experimental result showing the influence of the magnitude of $A_f$ is preserved.

### C. Local viscoelastic response of the polymer

At room temperature, the PS film polymers investigated are in the vitreous state and do have the same bulk elastic properties.[21–23] Therefore, the experimental results should normally be identical and independent of the molecular weight; besides, we did observe a large change in the friction behavior as a function of the molecular weight showing that the mechanical properties of the surface are different than that of the bulk.[16,21]

Let us assume that for polymer chains at the surface, a local viscoelastic response occurs when the intermediate and low molecular weights are studied. The main difficulty is to estimate the elastic modulus. Taking an elastic modulus at zero frequency $G(\omega=0)=10^8\ \text{N m}^{-2}$, with a protuberance of diameter $\Phi\sim 10$ nm, the stiffness of the protuberance scales as $k_0=G_0\Phi$ and $k_0=1\ \text{N m}^{-1}$. The mechanical susceptibility is given by

$$J_0=\frac{1}{G_0\Phi}. \tag{8}$$

Relaxation processes in polymer materials usually exhibit several relaxation times. Here, to simplify, we uniquely consider one relaxation time and the viscoelastic response is

$$J(t)=J_0[1-\exp(-t/\tau_N)], \tag{9}$$

where $J_0$ and $\tau_N$ are molecular weight dependent. The elastic displacement is given by

$$h(t)=\int_{-\infty}^{t} J(t-t')\dot{F}(t')dt'. \tag{10}$$

We further assume that the external force is given by the leading term of Eq. (7). That is, the static part of the right-hand side (r.h.s.) $F_{ext}\tau_{res}/T$. The action of the tip becomes a simple steplike function [Fig. 6(b)] and the initial time $t=0$ is the time at which the protuberance starts the growth process. With a step function, Eq. (10) gives an elastic displacement at time $t$:

$$h(t)=J_0F_{ext}\frac{\tau_{res}}{T}[1-\exp(-t/\tau_N)]. \tag{11}$$

However, even with this oversimplified approach, because $\tau_{res}$ varies as a function of the time, Eq. (11) cannot be used as is. After each cycle the height of the protuberance increases, consequently the amplitude of the oscillator decreases and the residence time $\tau_{res}$ increases. This makes the source term $J_0F_{ext}(\tau_{res})/T$ highly nonlinear and induces an avalanche effect. At the $k$th cycle, the residence time $\tau_{res}^k$ is given by the amplitude of the oscillator at the $(k-1)$th cycle. The amplitude is $A_f^{k-1}=A_f-h_{k-1}$, where $h_{k-1}$ is the height of the protuberance before the $k$th cycle. When the sample moves downwards the quantity of interest reducing the vibrating amplitude is not the height of the protuberance, but the vertical location $z_{k-1}$ of the sample with respect to the tip. Thus we subtract the vertical displacement of the piezoactuator which, at the $k$th cycle, is equal to the product $kv_pT$. Equation (6) is replaced by

$$\tau_{res}^k=\frac{T}{\pi}\cos^{-1}\left(1-\frac{d_c}{(A_f-z_{k-1})}\right), \tag{12}$$

where $z_{k-1}=h_{k-1}-kv_pT$ or, equivalently, $z_{k-1}=z_{k-2}+dh_{k-1}-v_pT$.

An attempt to reproduce the influence of the avalanche effect is to differentiate Eq. (11) to give the elastic displacement per oscillation:

$$dh_k=\frac{T}{\tau_N}J_0F_{ext}\frac{\tau_{res}^k}{T}\exp\left(-\frac{\tau_{res}^k}{\tau_N}\right). \tag{13}$$

Thus a height of the nanoprotuberance at the $k$th cycle $h_k=h_{k-1}+dh_k$ and a sample location given by

$$z_k=h_k-kv_pT. \tag{14}$$

The above numerical approach should be able to simulate the observed features for memory effects remaining negligible within the short time scale of the fast part of the amplitude variation.

Numerical results are reported in Fig. 7. The general trend is quite well reproduced. Quantitatively, it is not useful to separate the magnitude of $F_{ext}$ and $J_0$. The crude approximation of a rectangular periodic force together with our estimation of $J_0$ implies that it is not significant to separate the two terms. Thus, it is the product $J_0F_{ext}$, which is varied in the simulation.

The influence of the velocity at which the piezoactuator retracts is illustrated in Fig. 8. At $v_p=20$ nm/s, the minimum is slightly below the one observed at $v_p=400$ nm/s. Here again, the simulation gives a good qualitative picture.

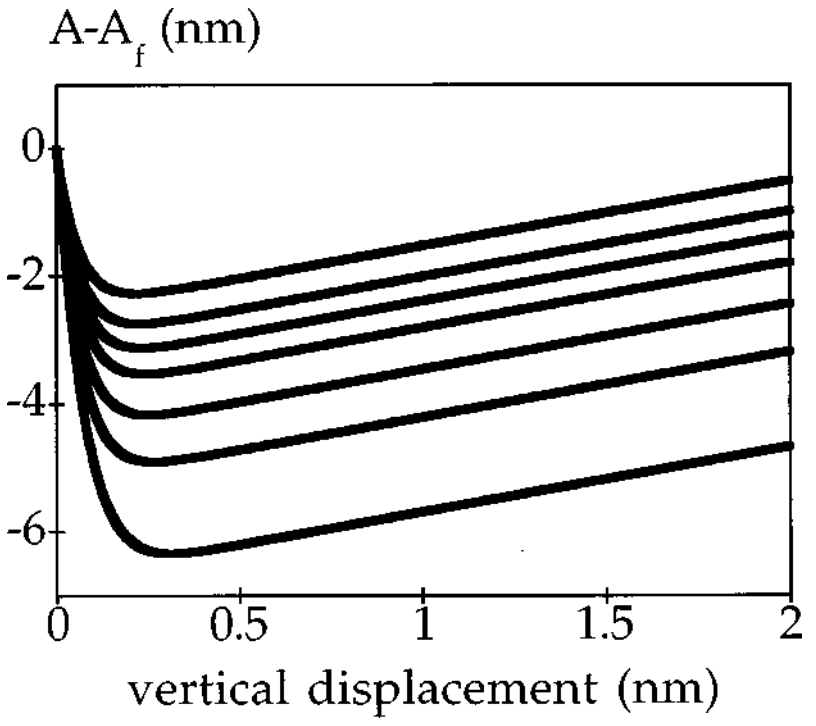

FIG. 7. Results of the numerical simulations using Eqs. (12), (13), and (14). The parameters are $J_0F_{ext}=15$ nm, $\tau_N=143\times10^{-6}$ s, and $d_c=3$ nm. From the lower to the upper curve the starting amplitudes are 8, 10, 12, 15, 18, 22, and 30 nm.

An attempt to fit the variation of the growth rate as a function of $A_f$ is given in Fig. 9. The general trend is quite well reproduced, but also is shown the inability of the simulation to quantitatively reproduce both the fast and slow rate regime. As discussed below (Sec. IV), one reason is that the product $J_0F_{ext}$ and the relaxation time $\beta$ are kept constant whatever the protuberance height.

## IV. DISCUSSION

The height dependence of the residence time does not enable us to solve analytically the growth process. Nevertheless, Eqs. (12) and (13) describing the fast part of the amplitude variations provide information about the relevant pa-rameters that allow the growth process to take place.

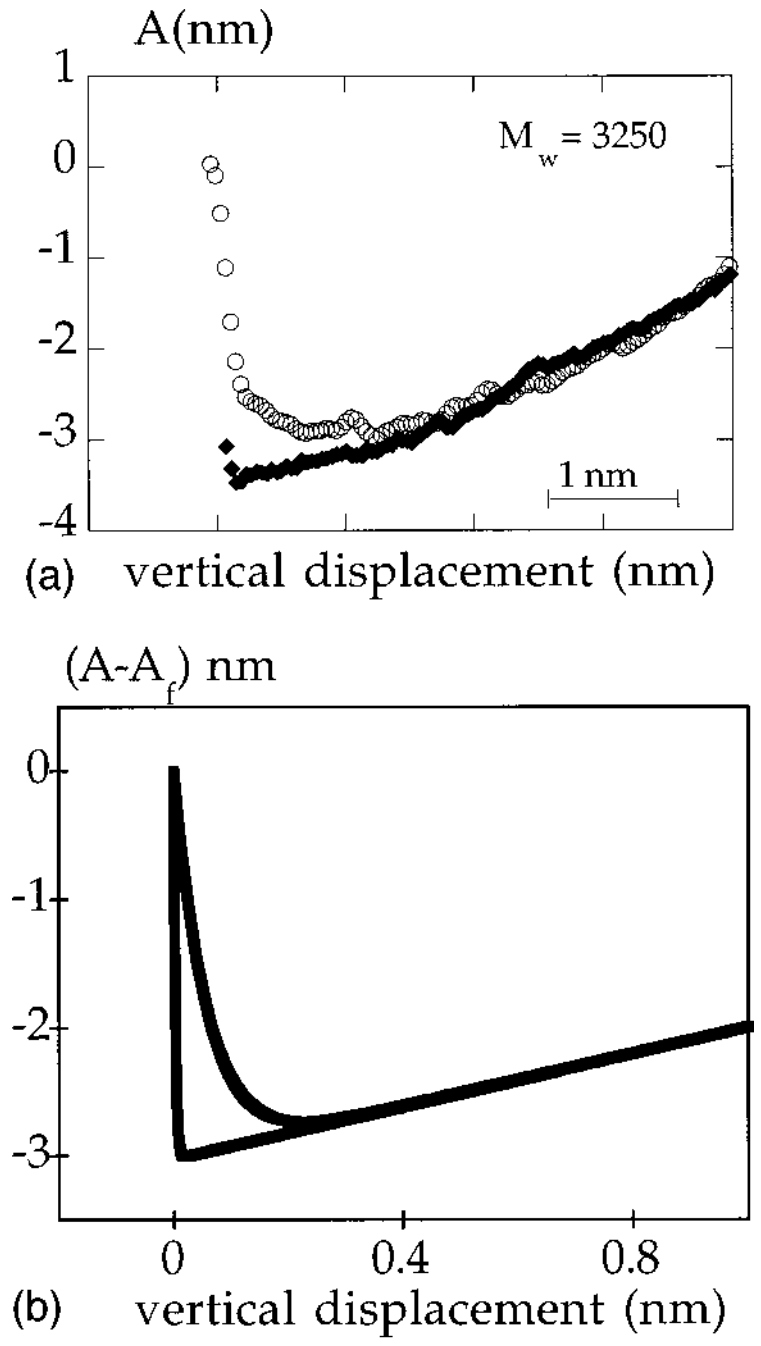

FIG. 8. Zoom of the fast part of the variation of the oscillating amplitude for two velocities of the piezoactuator. (a) Experimental data $M_w=3250$ and $A_f=20$ nm. Empty symbol $v_p=400$ nm/s, filled symbol $v_p=20$ nm/s. (b) Numerical simulation with the parameters $J_0F_{ext}=14$ nm, $\tau_N=143\times10^{-6}$ s, $d_c=3$ nm and the starting amplitude 15 nm.

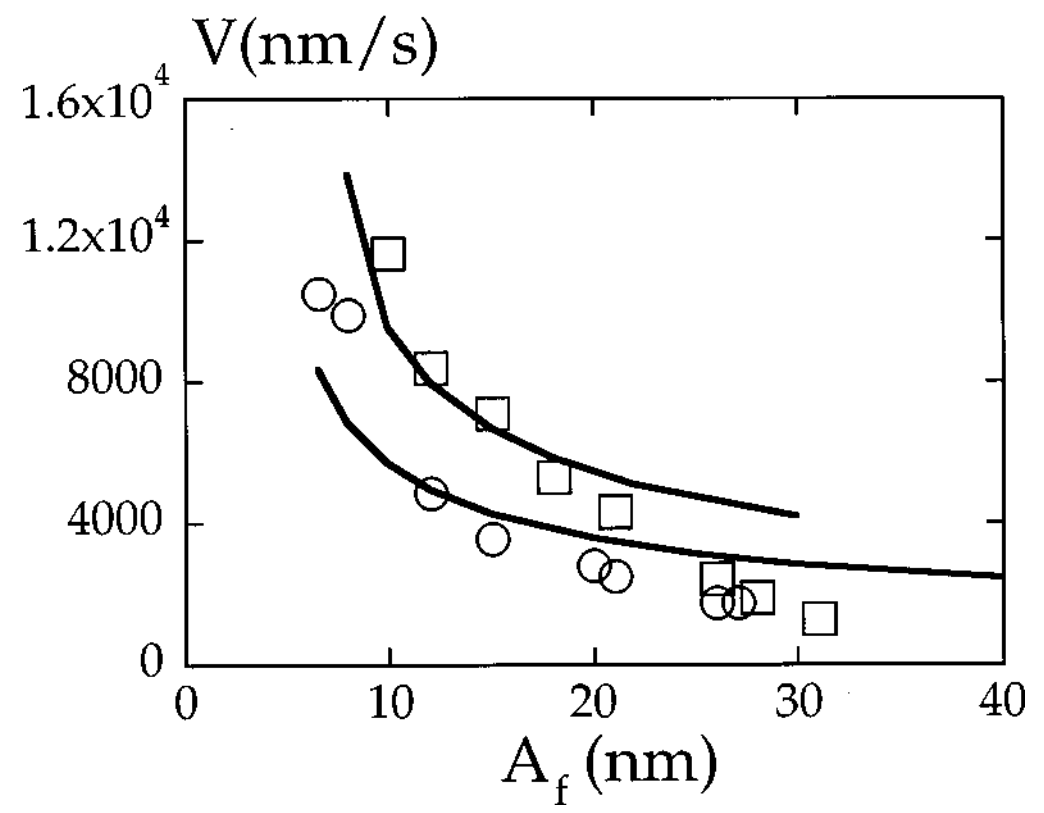

FIG. 9. Observed growth rates compared to the numerical results. Empty circle $M_w = 95\,000$, empty square $M_w = 23\,000$. The parameters are $J_0 F_{ext} = 20$ nm, $\tau_N = 143 \times 10^{-6}$ s, $d_c = 4$ nm, and $J_0 F_{ext} = 8$ nm, $\tau_N = 67 \times 10^{-6}$ s, $d_c = 3$ nm, respectively, for $M_w = 23\,000$ and $M_w = 95\,000$.

At this stage, it is worth to discuss the crude assumptions employed to describe the local mechanical susceptibility of the surface. The most interesting, but also the most difficult, is an appropriate estimation of this local response as many parameters as the protuberance shape and the effective force are unknown. One assumption concerns the choice of the analytical expression of the attractive force [Eq. (4)] here was used a sphere-plane surface interaction. A protuberance ended with an hemispheric shape with a radius $R$ can also be considered. In this case a sphere-sphere interaction must be used. Following the Derjaguin approximation,[15] the attractive interaction goes asymptotically towards a strength of interaction half that given by Eq. (4). As a consequence the attractive force becomes a function of the protuberance's height and decreases as the height increases. This geometrical effect has to be introduced with an expansion in term of $h/R$ in expression (4); including this effect does not make, however, any marked changes in the results presented above.

Another point is the local stiffness of the polymer. To estimate the local stiffness we need to estimate the size of the local surface interacting with the tip. The diameter $\Phi$ of this area is roughly defined by twice the radius of curvature of the tip. The local stiffness varies as $G\,\Phi$, thus typical values of the product $J_0 F_{ext}$, between 5 and 10 nm, with $R = 10$ nm and $F_{ext} = 2$ nN, give an elastic modulus $G$ between $10^8$ and $5 \times 10^7$ N m$^{-2}$. Here again, because of the change of the geometry of the sample, we might consider that the local stiffness is a function of the height of the protuberance. A simple structure, like a cylinder of diameter $\Phi$ and height $h$ (Fig. 5), provides an estimation of the influence of this geometrical effect. For a height large enough the stiffness $G\,\Phi$ must be replaced by a stiffness scaling as $G\Phi^2/h$. With $\Phi = 20$ nm, a change of the local stiffness becomes significant for heights larger than 20 nm. The maximum height deduced from the experimental data is found to be around 15 nm. Thus considering the geometrical factors as having a weak effect on the growth process is a reasonable assumption.

The third point is the intrinsic nonlinearities due to the locality of the mechanical response. The experimental results show two regimes in the kinetics, one very fast, particularly when low amplitudes are used, the second slower, when the amplitude returns to the $A_f$ value. The slow regime appears unambiguously after the minimum has been reached with a rather flat extremum. As demonstrated with the numerical results, the height dependence of the residence time clearly account for the fast part, but gives a poor agreement for the slow domain or just after the minimum. This could be due to change of the characteristic time $\tau_N$ and of the stiffness during the growth process of the nanoprotuberance. The picture drawn in Fig. 5 gives an oversimplified description of the nanoprotuberance properties. A more realistic picture should probably borrow ideas from models describing grazes in a vitreous polymer.[24,25] Because the polymer experiences large constraints at a local scale, the structure of the polymer can be different than that in the vitreous bulk state. This, indeed, is implicitly assumed when a viscoelastic response is introduced [Eq. (9)]. Also, nothing is said about the way the constraint is transmitted to the bulk. Therefore we may expect a nonlinear response coming from the intrinsic change of the polymer properties at the local scale.

The change of the relaxation time as a function of the height is difficult to assume. We might either consider an increase of the relaxation time as the size of the protuberance increases, which is the usual relation between size and time relaxation or, on the other hand, for a column structure one can expect the stability of the structure to be a decreasing function of the height. Therefore a relaxation time decreasing as the protuberance height increases.

To end this qualitative analysis, another possibility must be discussed to explain the unusual variation of the oscillating amplitude. A polymer neck could be created between the tip and the surface. Such a situation had been observed with the polysiloxane chains grafted on a silica surface.[26] In that case, using the static contact mode, the additional elastic force of entropic origin gives an additional deflection of the microlever. However, the neck of polymer was created because of the capability of the polysiloxane chains to be grafted at the apex of the tip. Such a process cannot occur with the PS chains. In addition, as shown in this paper, the elastic response was uniquely observed for polymers of molecular weight larger than that of the entanglement mass $M_e$. The entanglement mass of the PS chains is approximately 10 000 and the polymer films showing the most spectacular variations correspond to $M_w = 3250$ and $M_w = 1890$. Therefore an elastic contribution due to a neck of polymer between the tip and the film is quite unlikely to occur.

In this work were also shown situations in which the surface forces did have a significant contribution.[26] Evaluations were done showing that surface forces become significant when the elastic modulus are below $10^6$ N m$^{-2}$. Such values are much lower than that of the PS films ($10^9$ N m$^{-2}$) and the lowest elastic modulus estimated for the nanoprotuberances, around $10^7$ N m$^{-2}$. Nevertheless, at the very beginning, when the elastic deformation is small, one cannot completely avoid a possible contribution of the surface forces for the most sensitive samples.

To answer the complex questions concerning the intrinsic nonlinearity, it would be of great help to find experimental situations for which an analytical expression could be used. The sample of $M_w = 1890$ gives such an opportunity, the experimental results show a measurable variation of the am-plitude at large $A_f$, between 28 and 44 nm, with a protuberance height varying between 1.7 and 4 nm within a few milliseconds. Thus, the arccos function of the residence time can be expanded as a function of the ratio $z(t)/A_f$, giving the opportunity to make further assumptions to extract an analytical expression. Where $z(t)=h(t)-v_p t$ is the effective vertical location of the sample,

$$\frac{\tau_{\text{res}}}{T}=\frac{1}{\pi}\cos^{-1}\left(1-\frac{d_c}{A_f-[h(t)-v_p t]}\right)$$
$$=\frac{1}{\pi}\sqrt{\frac{2d}{A_f}}\left(1+\frac{1}{2}\frac{h(t)-v_p t}{A_f}\right)$$
$$+O\left(\frac{1}{\pi}\sqrt{\frac{2d}{A_f}}\left(1+\frac{3}{2}\frac{h(t)-v_p t}{A_f}\right)\right)^3. \qquad (15)$$

Assuming $\tau_{\text{res}}$ nearly constant throughout the experiment gives $\tau_{\text{res}}/T=(1/\pi)\sqrt{2d_c/A_f}$ and the elastic displacement is simply given by

$$h(t)=J_0F_{\text{ext}}\frac{1}{\pi}\sqrt{\frac{2d_c}{A_f}}[1-\exp(-\beta t)], \qquad (16)$$

taking into account the first order term of the Taylor expansion Eq. (15) gives a self consistent equation remaining analytically soluble (Appendix, Eq. (A7). The solution does have the same structure as the one given by the expression (16) [Appendix, Eqs. (A9) and (A11)] and have been used to fit the experimental data.

Not included in the above approach is taking into account the elastic contact between the tip and the sample. The effect of the contact between two elastic solids can be simply expressed through the stiffness ratio between the cantilever stiffness $k_c$ and the contact one $k_\phi=G\Phi$. In contact mode, a linear relationship between the cantilever deflection and the sample displacement is observed and the slope can be less than 1 if the contact stiffness is of the same order or less than the cantilever stiffness.[27] The slope $\alpha$ between the cantilever deflection and the piezoactuator displacement $Z_p$, $A=\alpha Z_p$, is given by $\alpha=1/[1+(k_c/k_\phi)]$. Therefore, a change of the cantilever deflection is given by $\alpha Z_p=\alpha v_p t$, and the observed velocity becomes $\alpha v_p$. For intermittent contact it is not obvious that such an expression is straightforwardly applied, particularly when dissipation happens during the contact between the tip and the sample. It is far beyond the scope of the present paper to discuss in details the usefulness of the above equation to describe the intermittent contact.[28] The equation to fit the experimental data is of the form

$$z(t)=S_e[1-\exp(-\beta_e t)]-\alpha v_p t, \qquad (17)$$

where $S_e$ and $\beta_e$ have the structure given in the Appendix.

Fits of the experimental variations are reported in the Fig. 10. Attempts to fit the results obtained for the sample of $M_w=23\,000$ and $M_w=3250$ at the highest amplitudes were also done. Since the experimental curves are correctly fitted with Eq. (17), it becomes possible to check the validity of the approximations used. For example, for free amplitudes $A_f$ varying between 32 and 44 nm, the effective vertical location $z(t)$ varies between 2.7 and 1.3 nm, corresponding to relative variations of the amplitude of 8.4% and 3%, respectively.

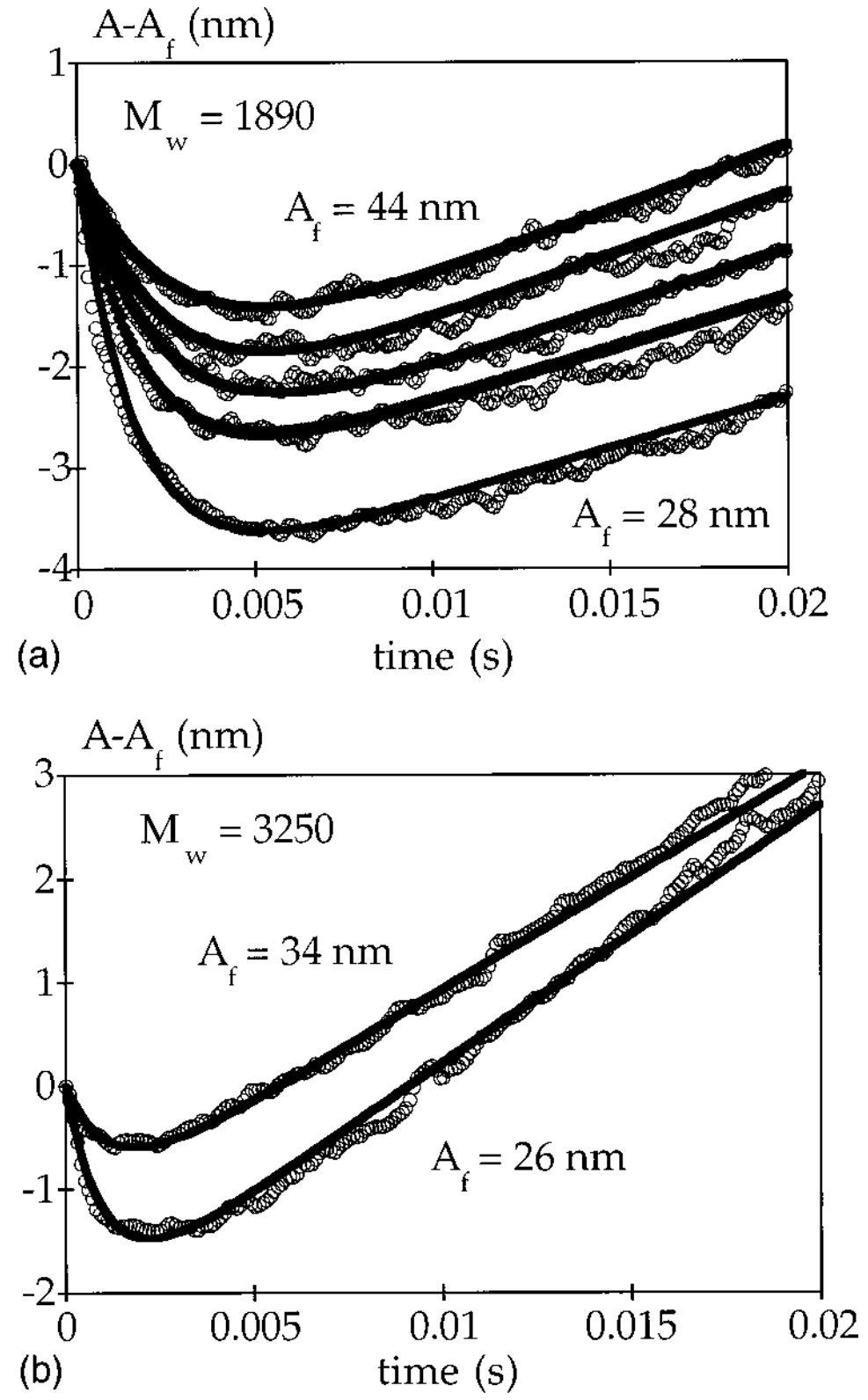

FIG. 10. Comparison between the experimental data and computed curves given by the Eq. (A11) (see Appendix).

Furthermore, since these corrections are partly taken into account with the first-order term of the expansion (15), one can expect that the fitted values of the product $J_0F_{\text{ext}}\sqrt{2d_c}$ and the relaxation time $\tau_N=\beta_{N^{-1}}$ are independent of $A_f$ and, in turn, of the protuberance height. As shown in Fig. 11, these quantities that aim to describe the intrinsic properties of the nanoprotuberance vary as a function of the free amplitude $A_f$. Therefore, the attempt to fit the experimental curves shows that Eq. (17) does have a too simple structure to describe satisfactorily the whole kinetics of the nanoprotuberance. The fact that these two quantities vary as a function of the height of the protuberance can have two origins: one is the change of the intrinsic properties as function of the protuberance size as discussed above, the second is due to the use of a simple viscoelastic response [Eq. (9)] assuming an average homogeneous response of the protuberance instead of an heterogeneous one as it occurs with a pointlike force applied on an elastic medium.[29]

The numerical simulation shows the origin of the purely experimental nonlinearity coming from the way the experiment is performed. For small protuberance height with respect to the free oscillating amplitude of the cantilever, an analytical solution can be derived that allows fits to be fruitfully compared to the experimental results. In most cases, a simple rheological model is unable to describe the whole growth process.

The above results show that by recording an image on a soft material with the tapping mode gives a topography that is a function of the experimental conditions employed. For example, the step height of a liquid film appears as a direct function of the free amplitude used. The smaller the free amplitude, the higher the observed height of the step.[30] Also, the height of islands or aggregates of molecules grafted on a surface may become a function of their mechanical susceptibility. Therefore, the image recorded becomes a mixing of a true topography and the mechanical response. Since the leading term of the action of the tip varies as $\sqrt{1/A_f}$, an easy way to check the origin of the contrast will be given by recording two images with two different free oscillating amplitudes.

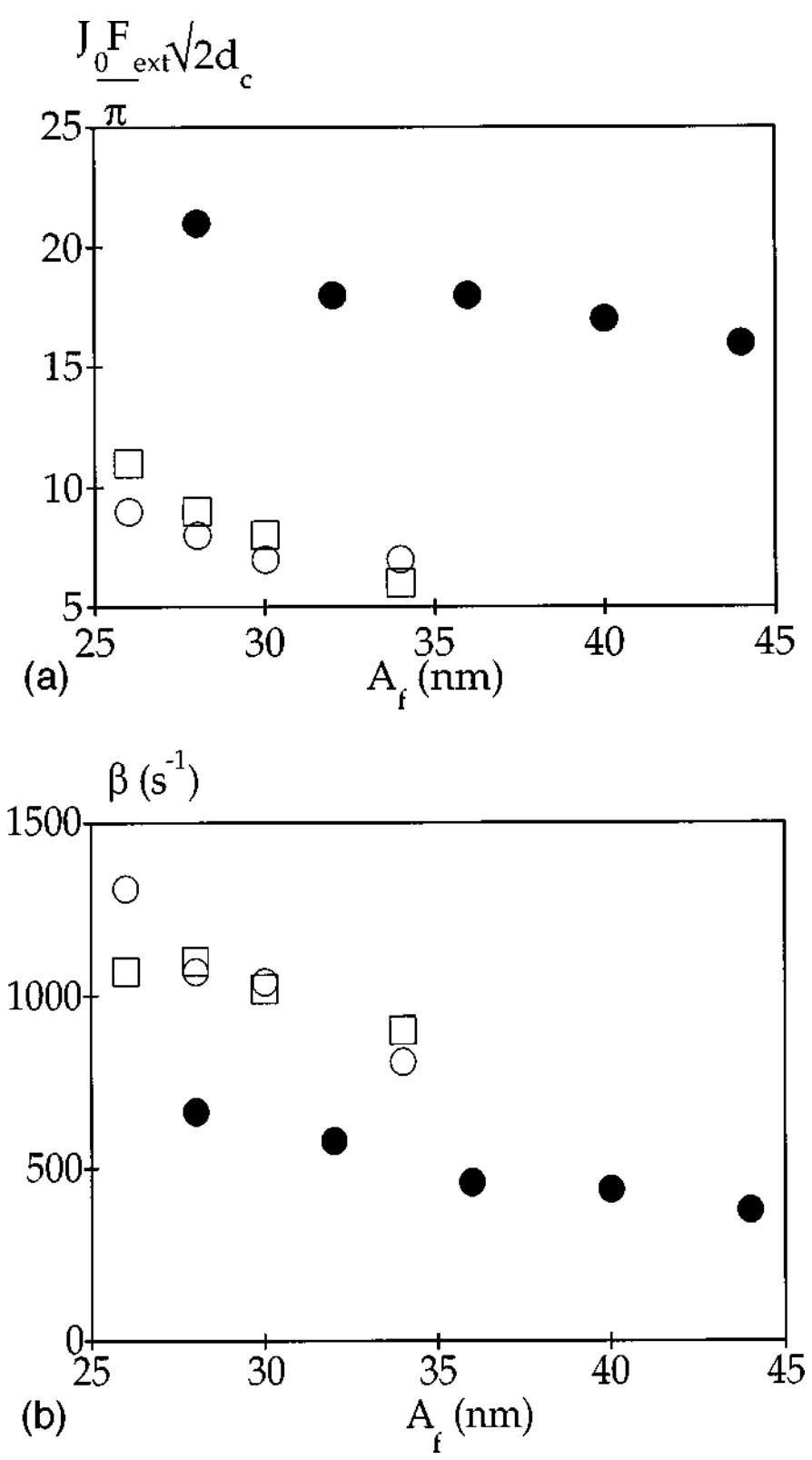

FIG. 11. Parameters $(J_0F_{\rm ext}/\pi)\sqrt{2d_c}$. (a) and $\beta$ (b) calculated from the fits shown in Fig. 10. The parameters $\alpha$ remain constant: $\alpha=0.25\,(2)$ $(M_w=1890)$, $\alpha=0.53$ (2) $(M_w=3250)$, $\alpha=0.50$ (1) $(M_w=23\,000)$. Filled circle, $M_w=1890$; empty circle, $M_w=3250$; empty square, $M_w=23\,000$.

## V. CONCLUSION

The aim of the present work was an attempt to investigate the ability of a vibrating tip-cantilever system, the Tapping mode, to probe mechanical properties without or only slightly touching the surface. On the one hand, to achieve the goal of minimizing the contact time between the tip and the surface, we use a criterion based on the study of the cycle of the hysteresis of the vibrating amplitude induced by an attractive field. On the other hand, polystyrene polymer films of various molecular weights were chosen as model samples exhibiting different mechanical properties at the surface. Anomalous variations of the oscillating amplitude were interpreted as a response of the vibrating lever to the growth of a nanoprotuberance. The smaller the molecular weight, the more sensitive is the polymer to the strength of the attractive field due to the proximity of the tip. In addition, the smaller the initial oscillating amplitude, the faster is the growth rate of the nanoprotuberance.

A numerical simulation shows the nonlinear relation between the oscillating amplitude and the growth rate, beside linear assumptions appear suitable enough to reproduce qualitatively the main features describing growth rates between 3000 and 15 000 nm/s. At high initial oscillating amplitudes, or equivalently short residence time of the tip near the surface, the growth process is found to be amenable to a simple analytical solution. The main interest of the analytical solution is to exhibit variation of the nanoprotuberance properties during the growth process that our present description is unable to describe. Further work is needed to investigate the nonlinearity of the mechanical response at the local scale.

The present works show unambiguously that mechanical properties can be probed by simply approaching a vibrating tip. It is also shown that the action of the tip can be simply described by considering uniquely the static component of the force. A direct consequence is that the strength of the attractive interaction can be finely tuned by varying the amplitude of the vibrating lever.

## APPENDIX

The static part of the force acting on the sample is given by Eq. (7):

$$F=F_{\rm ext}\frac{\tau_{\rm res}}{T} \tag{A1}$$

because the vibrating amplitude of the microlever varies as a function of the height of the protuberance, the static part has a time dependence through the relation

$$\begin{aligned}\frac{\tau_{\rm res}}{T}&=\frac{1}{\pi}\cos^{-1}\left(1-\frac{d_c}{A_f-(h(t)-v_pt)}\right)\\&=\frac{1}{\pi}\sqrt{\frac{2d_c}{A_f}}\left(1+\frac{1}{2}\frac{h(t)-v_pt}{A_f}\right)\\&\quad+O\left[\frac{1}{\pi}\sqrt{\frac{2d}{A_f}}\left(1+\frac{3}{2}\frac{h(t)-v_pt}{A_f}\right)\right]^3.\end{aligned} \tag{A2}$$

The height of the protuberance is given by

$$h(t)=\int_{-\infty}^{t}J(t-t')\dot{F}(t')dt'. \tag{A3}$$

Using the Laplace transform, we have

$$L\left(\int_{-\infty}^{t}J(t-t')\dot{F}(t')dt'\right)=pJ(p)F(p) \tag{A4}$$

and the following equation to solve

$$h(p)=pJ(p)F(p). \tag{A5}$$

Taking the first-order term of the Eq. (A2) gives

$$F(p)=\frac{F_{\rm ext}}{\pi}\sqrt{\frac{2d_c}{A_f}}\left(\frac{1}{p}+\frac{h(p)}{2A_f}-\frac{v_p}{2A_f}\frac{1}{p^2}\right); \tag{A6}$$

consequently the Laplace transform of the protuberance height is

—

$$h(p)=\frac{F_{\text{ext}}}{\pi}\sqrt{\frac{2d_c}{A_t}}\,J(p)\,\frac{\left(1-\frac{v_p}{2A_f}\frac{1}{p}\right)}{\left(1-\frac{\frac{F_{\text{ext}}}{\pi}\sqrt{\frac{2d_c}{A_f}}}{2A_f}\,pJ(p)\right)}. \tag{A7}$$

With a viscoelastic response of the form $J(t)=J_0\,[1-\exp(-\beta t)]$, the Laplace transform is

$$J(p)=\frac{J_0\beta}{p(p+\beta)}. \tag{A8}$$

Inserting Eq. (A8) into (A7) and solving the inverse Laplace transform gives the time dependence of the protuberance height:

$$h(t)=S_e[1-\exp(-\beta_e t)]-\frac{J_0F_{\text{ext}}}{\pi 2A_f}\sqrt{\frac{2d_c}{A_f}}\frac{\beta}{\beta_e}\,v_p t \tag{A9}$$

with

$$\beta_e=\beta\left(1-\frac{J_0F_{\text{ext}}}{\pi 2A_f}\sqrt{\frac{2d_c}{A_f}}\right),$$

$$S_e=\frac{J_0F_{\text{ext}}}{\pi}\sqrt{\frac{2d_c}{A_f}}\frac{\beta}{\beta_e}\left(1+\frac{v_p}{2A_f\beta_e}\right);$$

thus a vertical location of the sample is given by

$$\begin{aligned}z(t)&=h(t)-v_p t\\ &=S_e[1-\exp(-\beta_e t)]-\left(1+\frac{J_0F_{\text{ext}}}{\pi 2A_f}\sqrt{\frac{2d}{A_f}}\frac{\beta}{\beta_e}\right)v_p t.\end{aligned} \tag{A10}$$

The final equation used to fit the experimental data is

$$\begin{aligned}z(t)&=h(t)-\alpha v_p t\\ &=S_e[1-\exp(-\beta_e t)]-\alpha\left(1+\frac{J_0F_{\text{ext}}}{\pi 2A_f}\sqrt{\frac{2d_c}{A_f}}\frac{\beta}{\beta_e}\right)v_p t\end{aligned} \tag{A11}$$

and the parameters fitted are the product $(J_0F_{\text{ext}}/\pi)\sqrt{2d_c}$, $\beta$, and $\alpha$.

---

*FAX: 33 5 56 84 69 70.
Electronic address: jpaime@frbdx11.cribx1.u-bordeaux.fr

[1] T. R. Albrecht, P. Grütter, D. Horne, and D. Rugar, J. Appl. Phys. **69**, 668 (1991).
[2] Y. Martin, C. C. Williams, and H. K. Wickramasinghe, J. Appl. Phys. **61**, 4723 (1987).
[3] G. M. Mc Clelland, R. Erlandsson, and S. Chiang, Rev. Prog. Quant. Nondestr. Eval. **6**, 1307 (1987).
[4] B. D. Terris, J. E. Stern, D. Rugar, and H. J. Mamin, Phys. Rev. Lett. **63**, 2669 (1989).
[5] S. Hudlet, M. SaintJean, B. Roulet, J. Berger, and C. Guthman, J. Appl. Phys. **77**, 3308 (1995).
[6] F. J. Giessibl, Science **267**, 68 (1995).
[7] Y. Sugarawa, M. Otha, H. Ueyama, and S. Morita, Science **270**, 1646 (1995).
[8] D. Anselmetti, R. Lüthi, E. Meyer, T. Richmond, M. Dreier, J. E. Frommer, and H. J. Güntherodt, Nanotechnology **5**, 87 (1994).
[9] S. Kitamura and M. Iwatsuki, Jpn. J. Appl. Phys., Part 2 **35**, L668 (1996).
[10] P. Gleyzes, P. K. Kuo, and A. C. Boccara, Appl. Phys. Lett. **58**, 2989 (1991).
[11] B. Anczykowski, D. Krüger, and H. Fuchs, Phys. Rev. B **53**, 15 485 (1996).
[12] J. de Weger, D. Binks, J. Moleriaar, and W. Water, Phys. Rev. Lett. **76**, 3951 (1996).
[13] R. Boisgard, D. Michel, and J. P. Aimé, Surf. Sci. **401**, 199 (1998).
[14] Digital Instruments (Santa Barbara).
[15] J. Israelachvili, *Intermolecular and Surface Forces* (Academic, New York, 1992).
[16] D. Michel, S. Marsaudon, and J. P. Aimé, Tribol. Lett. **4**, 75 (1998).
[17] S. Gauthier, J. P. Aimé, T. Bouhacina, A. J. Attias, and B. Desbat, Langmuir **12**, 5126 (1996).
[18] R. G. Winkler, J. P. Spatz, S. Seiko, M. Möller, and O. Marti, Phys. Rev. B **54**, 8908 (1996).
[19] J. Tamayo and R. Garcia, Langmuir **12**, 4430 (1996).
[20] S. N. Magonov, V. Elings, and M. H. Whangbo, Surf. Sci. **389**, 201 (1997).
[21] D. Michel, thesis, Université Bordeaux I (Talence, France), 1997.
[22] J. Perez, *Physique et Mécanique des Polymères Amorphes* (Tec-Doc, Paris, 1991).
[23] T. G. Fox and P. J. Flory, J. Appl. Phys. **21**, 581 (1950).
[24] E. Kramer and L. Berger, Adv. Polym. Sci. **91**, 1 (1990).
[25] P. G. de Gennes, Europhys. Lett. **15**, 191 (1991).
[26] J. P. Aimé and S. Gauthier (unpublished).
[27] J. P. Aimé, Z. Elkaakour, C. Odin, T. Bouhacina, D. Michel, J. Curély, and A. Dautant, J. Appl. Phys. **76**, 754 (1994).
[28] A numerical simulation similar to the ones given in Refs. 18, 19, and 21, in which a contact area computed with the Hertz model is used, gives a slope of 0.16 for an elastic modulus of $10^7$ N m$^{-2}$ and a cantilever stiffness $k_c=3.5$ N m$^{-1}$. Note that using the same values gives a slope of 0.04 for the contact deflection mode. L. Nony, R. Boisgard, and J. P. Aimé (unpublished).
[29] L. Landau and E. Lifshitz, *Theory of Elasticity* (Mir, Moscow, 1967).
[30] S. Bardon (unpublished).